# The CTLNet for Shanghai Composite Index Prediction

Haibin Jiao[1]

[1]Independent Researcher
jiaohaibin@alu.ruc.edu.cn

**Abstract.** Shanghai Composite Index prediction has become a hot issue for many investors and academic researchers. Deep learning models are widely applied in multivariate time series forecasting, including recurrent neural networks (RNN), convolutional neural networks (CNN), and transformers. Specially, The Transformer encoder, with its unique attention mechanism and parallel processing capabilities, has become an important tool in time series prediction, and has an advantage in dealing with long sequence dependencies and multivariate data correlations. Drawing on the strengths of various models, we propose the CNN-Transformrer-LSTM Networks (CTLNet). This paper explores the application of CTLNet for Shanghai Composite Index prediction and the comparative experiments show that the proposed model outperforms state-of-the-art baselines.

## 1 Introduction

The Shanghai Composite Index, as one of the most representative and influential indices in China's securities market, is an indispensable reference for studying and judging the trend of stock price changes. The prediction of the Shanghai Composite Index has become a hot topic. Deep learning models are widely applied in multivariate time series forecasting and have achieved great success. To understand complex patterns and their relationships, many methods have emerged, including CNN [1, 2], LSTM [3, 4, 5], transformers [6, 7，8，9], and Graph Neural Networks (GNN) [10, 11]. Some works[12] proposes a novel Convolutional Long Short-Term Memory network (ConvLSTM) for nowcasting precipitation, which combines the advantages of CNN and LSTM. Particularly, many Transformer-based methods have demonstrated remarkable performance, due to its unique attention mechanism and parallel processing capabilities in sequences. These methods mainly include the Temporal Transformer, which evolves from applying attention between time points to applying attention between sub-series. In addition, by representing time points or sub-series as nodes and their relationships as edges, GNNs can leverage the graph structure to capture intricate dependencies in the data. However, there are challenges that limit the effectiveness of these methods in multivariate time series forecasting, which can be summarized as follows.

For one thing, the temporal patterns and their relationships of time series are crucial for accurate predictions. However, individual time point in time series contains less information[13]. To address the issue of information sparsity, some works employ patch-based approaches [7] to enhance locality and some works use hypergraph structures [14] to capture temporal pattern interactions. However, these methods fail to make full use of the temporal information contained in the data. For another, financial markets are characterized by high volatility and complexity. Traditional financial forecasting methods often struggle to capture the intricate patterns and trends within these markets. The CNN-LSTM have shown promise in addressing these challenges and have emerged as a powerful tool for time series forecasting, due to its ability to extract local features (via CNN) and temporal dependencies (via LSTM). However, applying attention mechanisms to time series has become mainstream, which can promote more comprehensive pattern interactions.

Drawing on the strengths of various models, we propose the CTLNet, a comprehensive model which integrates the advantages of many models. Specifically, the CNN module functions as a local feature extractor, and we use the CNN module to aggregate adjacent time points into time nodes. We provide the sequence of nodes embeddings as an input to the Transformer encoder. Time nodes are treated the same way as tokens in an NLP application or as patches in Vision Transformer (ViT). The Transformer encoder module is introduced to promote comprehensive pattern interactions and functions as a global feature extractor. Finally, by taking all nodes in sequence as a time series, we use the LSTM module to capture temporal dependencies in time nodes, which is used by the output layer to make predictions. We train the model on time series regression forecasting in supervised fashion and the experiments show that the CTLNet achieves the best performance. The main contributions of this work are as follows:

- In order to enhance the locality of the input sequence and obtain its complex temporal patterns, The CNN module functions as a local feature extractor to obtain the local representation of the input sequence. Specially, The outputs of the CNN module, a 1D sequence of vectors, is regarded as the node embeddings, and each vector represents a time node.

- The Transformer encoder module does comprehensive interactions among time nodes and functions as a global feature extractor to obtain global representations of time nodes, thus enriching the interaction information of each time node and dressing the issue of information sparsity further.

- By taking all time nodes in sequence as a time series, The LSTM module functions as a feature extractor to capture temporal dependencies in time nodes.

- The designed CTLNet is not only a comprehensive model which integrates the advantages of many models and can be viewed as a variant or extension of many models, but also a concise and elegant model whose principles can be explained clearly. We perform experiments on 2 real-world datasets, and the experimental results demonstrate that it outperforms the state-of-the-art baselines.

## 2 Related Work

**The CNN-LSTM model.** The CNN-LSTM model, as a deep learning model that combines the advantages of CNN and LSTM, demonstrates strong application potential in financial market trend analysis, time series forecasting, and other fields. Some works[12] proposes a novel Convolutional Long Short-Term Memory network (ConvLSTM) for nowcasting precipitation, demonstrating the application potential of the model in the field of meteorological forecasting. Some study[15] provides a detailed introduction to the application of various deep learning methods in time series forecasting, including the analysis and comparison of hybrid models. such as CNN-LSTM.

**Transformer-based models.** The initial Transformer-based models refer to time points[16,17,18] as tokens and apply attention mechanism between them. Subsequent works demonstrated that attention mechanism at a sub-series level is more effective and can reduce the complexity due to fewer tokens. PatchTST [7] provides a general paradigm of the Temporal Transformer at the sub-series level.

**Graph Neural Network.** When dealing with complex relationships, traditional time series analysis methods may have difficulties in capturing the correlations among variables in the time dimension. The emergence of GNN brings hope to this predicament. GNN maps time points and variables into nodes, represents their correlations as edges, and effectively models time series data through graph structures, precisely learning complex pattern dependencies. Some works proposes a general GNN framework specifically designed for multivariate time series data, which is used to capture the potential spatial dependencies between variables. Some works [19, 20] study time series forecasting in the traffic domain with various GNN [21, 22, 23]. Recent studies [24, 25] show that hypergraph neural networks(HGNNs) have broad prospects in practical applications.

Taking into account the strengths of various models, we propose the CTLNet, a comprehensive model which can be viewed as a variant or extension of many models. The CTLNet mainly consists of three modules. The CNN module functions as a local feature extractor. The Transformer encoder module functions as a global feature extractor. The LSTM module functions as a feature extractor to capture temporal dependencies.

## 3 CNN-Transformrer-LSTM Networks

As mentioned above, the proposed CTLNet is a comprehensive model which integrates the advantages of many models. It not only addresses the issue of information sparsity by enhancing locality of time points and facilitating comprehensive pattern interaction**s**, but also makes full use of the temporal information contained in the data. Notably, the outputs of the CNN module is a 1D sequence of vectors, with each vector serving as a representation of a sub-series patch within the overall data. When we apply Transformer-Encoder to sub-series patches, the CTLNet becomes a type of PatchTST. In addition, through the CNN module, time series data can be represented as a sequence of time nodes, a type of graph data. Graph data not only have nodes, which can be represented as a vector, but also have information about the edges. In fact, as we apply self-attention of Transformer encoder to a sequence of time nodes, the CTLNet becomes a type of GNN, which can automatically generates the incidence matrices of nodes in the graph. To sum up, the CTLNet integrates three modules. It is a concise and elegant model, and its principles can be elaborated in detail. We first use the CNN module to aggregate adjacent time points into time nodes which are the local representations or the temporal patterns of time series. Then, the Transformer encoder module is introduced to facilitate comprehensive interactions among time nodes and functions as a global feature extractor to obtain global representations of time nodes. Finally, the LSTM module is introduced to capture temporal dependencies in time nodes by taking all time nodes in sequence. The output of the LSTM module is utilized by the output layer for making predictions. The overall framework of the network is shown in Figure 1.

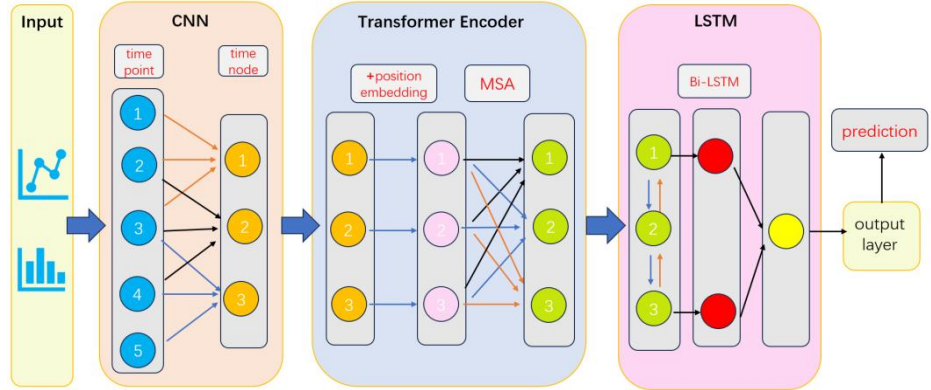

**Fig. 1.** The framework of CTLNet

### 3.1 CNN Module

CNN1D can automatically extract local features from time series, capture patterns, and perform prediction and classification. These characteristics make CNN1D have a wide range of application prospects in the field of time series analysis. The input sequence $x \in \mathbb{R}^{N \times D}$ is 5 time points in sequence. Through the sliding window mechanism, the CNN module aggregates 3 time points into a time node and can extract local features from 5 time points, thus laying foundations for interactions among time nodes. The network start by applying the following operations:

Processing the input sequence x ∈ $\mathbb{R}^{N \times D}$ with a convolutional module to obtain the all nodes in sequence z given by

$$z = \text{Conv}(x) \qquad (1)$$

where Conv is the 1D convolution, and z denotes the result of the aggregation function, and is a 1D sequence of nodes. Each node is represented by a vector as node embeddings.

**3.2 Transformer Encoder Module**

The standard Transformer receives as input a 1D sequence of token embeddings. The outputs of the CNN module is a 1D sequence of vectors. We refer to each vector as a node embedding. The position embeddings posE are added to the node embeddings z to retain positional information. We use standard learnable 1D position embeddings.

$$n = z + posE \qquad (2)$$

The resulting sequence of embedding vectors n serves as input to the Transformer encoder. The Transformer encoder consists of alternating layers of multi-headed self-attention (MSA) and MLP blocks. Layernorm (LN) is applied before every block, and residual connections after every block. We use MSA to promote pattern interactions and to enrich the interaction information of each time node. It can be formally expressed as follows:

$$m = \text{MSA}(\text{LN}(n)) + n \qquad (3)$$

$$l = \text{MLP}(\text{LN}(m)) + m \qquad (4)$$

**3.3 LSTM Module**

The LSTM Module processes the sequence nodes and captures long-term dependencies. LSTM units consist of three gates: the forget gate, the input gate, and the output gate. These gates control the flow of information through the network, enabling it to learn long-term dependencies and forget irrelevant information. The LSTM module models the temporal dependencies in the time nodes by updating its internal state at each time step based on the current input and the previous hidden state. This is particularly useful in financial markets where noise and volatility can obscure underlying trends.

After processing the input time nodes in sequence, the Bidirectional LSTM (Bi-LSTM) layer generates the feature representations v, which capture the temporal dependencies within the data and are used by the output layer to make predictions.

$$v = \text{BiLSTM}(l) \qquad (5)$$

### 3.4 A Prediction Module & Loss Function

After extracting features of all nodes by LSTM module, we feed them into a output layer for prediction. We choose Mean Squared Error (MSE) as our forecasting loss.

$$y = Output(v) \qquad (6)$$

## 4 Experiment

### 4.1 Setup

**Datasets.** For time series forecasting, we conduct experiments on 2 real-world datasets, including SSE Composite Index and SSE 50 Index. In the datasets, each data sample is collected at a frequency of 5 minutes, including the opening price, closing price, highest price, lowest price and trading volume. By adopting the sliding window method, we generate the training set and the test set from the datasets. We normalize both the training set and the test set to the range of 0 to 1. By using the five time points to predict the sixth time point, we train the model on time series regression forecasting in supervised fashion. Table 1 gives the datasets statistics.

**Table 1.** Datasets statistics.

| Datasets | Frequency | Sample size | The dimensionality of an individual sample |
| --- | --- | --- | --- |
| SSE Composite Index | 5m | 11616 | 6 |
| SSE 50 Index | 5m | 11616 | 6 |

**Baselines.** We compare CTLNet with 4 competitive baselines, i.e., LSTM, CNN-LSTM, the Transformrer-CNN-LSTM Networks (TCLNet) , and PatchTST.

**Metrics.** MAE are used as evaluation metrics and lower values mean better performance. $R^2$ is a core indicator used to quantify the explanatory power of a regression model, with its value ranging from 0 to 1. An $R^2$ closer to 1 means better performance.

**Experimental Settings**. SGD is set as the optimizer with the learning rate of 0.01 and the momentum 0f 0.9.

### 4.2 Training Results

We train all models, including CTLNet, LSTM, CNN-LSTM, and TCLNet. Table 2 shows the results.

Table 2. Training Results on SSE Composite Index datasets.

| Models | CTLNet | LSTM | CNN-LSTM | TCLNet |
|---|---|---|---|---|
| Parameters | 102,153 | **36,481** | 67,393 | 67,887 |
| Train Time | 97.58s | 38.23s | **37.17s** | 73.77s |
| Train MAE | 0.0079 | 0.0056 | **0.0053** | 0.0153 |

### 4.3 Ablation Studies and Overall Comparison

To investigate the effectiveness of this method, We first compare LSTM model to CNN-LSTM model. The experimental results on 2 datasets are shown in Table 3. As can be seen from Table 2 and Table 3, the CNN-LSTM model outperforms LSTM model on the SSE 50 Index dataset , while requiring less the training time, which shows the importance of extracting local features from time series and capturing patterns of time points. In addition, The CNN can effectively extract and compress the key information in the sequence, thus reducing the computational cost. This ability is crucial for enhancing the efficiency of predictions, especially when dealing with long sequence financial data. By comparing the CTLNet to CNN-LSTM model, we can observe that the CTLNet outperforms CNN-LSTM model on all datasets, showing the effectiveness of the Transformrer encoder module. It functions as a global feature extractor to obtain global representations of time nodes, thus enriching the interaction information of each time node. Specifically, the sequence of the TCLNet modules is: Transformrer encoder module, CNN module, LSTM module. When we placed the Transformrer encoder module at the very beginning, the TCLNet gets the worst forecasting results, which demonstrates the original time series of time points contains more noise interference and less effective information, compared to time nodes in sequence.

Table 3. Results of time series forecasting.

| Models | CTLNET | | LSTM | | CNN-LSTM | | TCLNET | |
|---|---|---|---|---|---|---|---|---|
| Metric | MAE | $R^2$ | MAE | $R^2$ | MAE | $R^2$ | MAE | $R^2$ |
| SSE Composite Index | **0.0108** | **0.971** | 0.0132 | 0.962 | 0.0138 | 0.959 | 0.0156 | 0.935 |
| SSE Composite Index50 | **0.0083** | **0.971** | 0.0101 | 0.962 | 0.0084 | 0.970 | 0.0154 | 0.906 |

## 5 Conclusions and Future Work

In this paper, we have explored the application of Transformers to time series. We interpret time series as a sequence of nodes obtained by the CNN module and process it by a standard Transformer encoder as used in NLP to capture interactions among nodes, and then we use LSTM Module processes the sequence nodes to captures temporal dependencies. We proposed CTLNet for the time series prediction of the SSE Composite Index, which can capture local feature in time series, promote comprehensive pattern interactions, and make full use of the temporal information contained in the data. Experimentally, The CTLNet is a good at addressing the problem of information sparsity and achieves the best performance.

While these initial results are encouraging, many challenges remain. For one thing, altering the size of the sliding window in CNN has a significant impact on time series forecasting, which requiring a trade-off between feature extraction capability, computational efficiency, and forecasting performance. For the same time series, when the CNN modules of the CTLNet employ different sliding window sizes, we can use the CTLNet as a whole to obtain different feature vectors, and then performs feature fusion. But it may result in an increase in computational complexity. Notably, when processing long input sequences, we can stack the CNN modules and the Transformrer encoder module as a whole to continuously compress the length of the sequence, thus reducing the computational cost. For another, time series data can be regarded as a sequence of time nodes, thus represented in the form of graph data. Specially, when we apply MSA to graphs data, it essentially becomes a type of GNN, which can automatically generates the incidence matrices of nodes in the graph. By considering the topological structure of time series, we may utilize the GNN for prediction. Thus, the challenge is how to combine the advantages of GNN in modeling complex structural information with the strengths of deep learning models in feature extraction.

## References

1. Shaojie Bai, J Zico Kolter, and Vladlen Koltun. An empirical evaluation of generic convolutional and recurrent networks for sequence modeling. arXiv preprint arXiv:1803.01271(2018).
2. Rajat Sen, Hsiang-Fu Yu, and Inderjit S Dhillon. Think globally, act locally: A deep neural network approach to high-dimensional time series forecasting. Advances in neural information processing systems, 32(2019).
3. Donghui Chen, Ling Chen, Youdong Zhang, Bo Wen, and Chenghu Yang. A multiscale interactive recurrent network for time-series forecasting. IEEE Transactions on Cybernetics,52(9):8793–8803(2021).
4. Ling Chen and Jiahua Cui. TPRNN: A top-down pyramidal recurrent neural network for time series forecasting. arXiv preprint arXiv:2312.06328(2023).
5. Zipeng Chen, Qianli Ma, and Zhenxi Lin. Time-aware multi-scale RNNs for time series modeling. In IJCAI, pages 2285–2291( 2021).
6. Tian Zhou, Ziqing Ma, Qingsong Wen, Xue Wang, Liang Sun, and Rong Jin. FEDformer: Frequency enhanced decomposed transformer for long-term series forecasting. In Proceedings of the International Conference on Machine Learning, pages 27268 – 27286(2022).
7. Yuqi Nie, Nam H Nguyen, Phanwadee Sinthong, and Jayant Kalagnanam. A time series is worth 64 words:Long-term forecasting with transformers. arXiv preprint arXiv:2211.14730(2022).
8. Yunhao Zhang and Junchi Yan. Crossformer: Transformer utilizing cross-dimension dependency for multivariatetime series forecasting. In The Eleventh International Conference on Learning Representations(2022).

9. Yong Liu, Tengge Hu, Haoran Zhang, Haixu Wu, Shiyu Wang, Lintao Ma, and Mingsheng Long. itransformer: Inverted transformers are effective for time series forecasting. arXiv preprint arXiv:2310.06625(2023).
10. Wanlin Cai, Yuxuan Liang, Xianggen Liu, Jianshuai Feng, and Yuankai Wu. MSGNet: Learning multi-scale inter-series correlations for multivariate time series forecasting. In Proceedings of the AAAI Conference on Artificial Intelligence, volume 38, pages 11141–11149(2024).
11. Ling Chen, Donghui Chen, Zongjiang Shang, Binqing Wu, Cen Zheng, Bo Wen, and Wei Zhang. Multi-scale adaptive graph neural network for multivariate time series forecasting. IEEE Transactions on Knowledge and Data Engineering, pages 10748–10761(2023).
12. Xingjian Shi, Zhourong Chen, Hao Wang, et al. "Convolutional LSTM Network: A Machine Learning Approach for Precipitation Nowcasting"[J]. Advances in Neural Information Processing Systems, 28:802-810(2015).
13. Haizhou Cao, Zhenhao Huang, Tiechui Yao, Jue Wang, Hui He, and Yangang Wang. InParformer: Evolutionary decomposition transformers with interactive parallel attention for long-term time series forecasting. In Proceedings of the AAAI Conference on Artificial Intelligence, volume 37, pages 6906–6915(2023).
14. Zongjiang Shang and Ling Chen. MSHyper: Multi-scale hypergraph transformer for long-range time series forecasting. arXiv preprint arXiv:2401.09261(2024).
15. Bryan Lim, Stefan Zohren. "Time-series forecasting with deep learning: a survey"[J]. arXiv preprint arXiv:2009.05407(2020).
16. Haoyi Zhou, Shanghang Zhang, Jieqi Peng, Shuai Zhang, Jianxin Li, Hui Xiong, and Wancai Zhang. Informer:Beyond efficient transformer for long sequence time-series forecasting. In Proceedings of the AAAI conference on artificial intelligence, volume 35, pages 11106–11115(2021).
17. Shiyang Li, Xiaoyong Jin, Yao Xuan, Xiyou Zhou, Wenhu Chen, Yu-Xiang Wang, and Xifeng Yan. Enhancing the locality and breaking the memory bottleneck of transformer on time series forecasting. Advances in neural information processing systems, 32(2019).
18. Shizhan Liu, Hang Yu, Cong Liao, Jianguo Li, Weiyao Lin, Alex X Liu, and Schahram Dustar. Pyraformer:Low-complexity pyramidal attention for long-range time series modeling and forecasting. In International conference on learning representations(2021).
19. Yaguang Li, Rose Yu, Cyrus Shahabi, and Yan Liu. Diffusion Convolutional Recurrent Neural Network: Data-Driven Traffic Forecasting. In ICLR(2018).
20. Zonghan Wu, Shirui Pan, Guodong Long, Jing Jiang, and Chengqi Zhang. Graph WaveNet for Deep Spatial-Temporal Graph Modeling. In IJCAI. 1907–1913(2019).
21. William L. Hamilton, Zhitao Ying, and Jure Leskovec. Inductive Representation Learning on Large Graphs. In NeurIPS. 1024–1034(2019).
22. Thomas N. Kipf and Max Welling. Semi-Supervised Classification with Graph Convolutional Networks. In ICLR(2017).
23. Kai Zhao, Yukun Zheng, Tao Zhuang, Xiang Li, and Xiaoyi Zeng. Joint Learning of E-commerce Search and Recommendation with a Unified Graph Neural Network. In WSDM. 1461–1469(2022).
24. Song Bai, Feihu Zhang, and Philip HS Torr. Hypergraph convolution and hypergraph attention. Pattern Recognition, 110:107637(2021).
25. Yichao Yan, Jie Qin, Jiaxin Chen, Li Liu, Fan Zhu, Ying Tai, and Ling Shao. Learning multi granular hypergraphs for video-based person re-identification. In Proceedings of the IEEE/CVF conference on computer vision and pattern recognition, pages 2899 – 2908(2020).